\documentclass{appolb}
\usepackage{amsmath}
\usepackage{graphicx}
\def\be{\begin{equation}}
\def\ee{\end{equation}}
\begin{document}
\title{Fluctuations in small systems
  \thanks{Presented at Diffraction and Low-x 2018, Reggio Calabria, August 2018}%
}%
\author{St\'ephane MUNIER
  \address{Centre de physique th\'eorique (CPHT),
    \'Ecole polytechnique, CNRS,\\ Universit\'e Paris-Saclay, Route de Saclay, 91128 Palaiseau, France}
}
\maketitle
\begin{abstract}
  We review the main features of event-by-event fluctuations of the content of the Fock
  states of onia (as models for dilute hadrons, or as bare
  hadronic components of virtual photons),
  as well as some of their observable consequences. We
  briefly address the total scattering cross section of a small onium off a nucleus,
  then of two small onia. Finally, we explain that the multiplicity in the final state
  of collisions of large onia with nuclei may directly be related to
  the gluon density in the former. We provide first predictions for the event-by-event
  fluctuations of the gluon density.
  \end{abstract}
%\PACS{PACS numbers come here}

\section{Dressed hadronic states from a branching process}

Consider an asymptotic onium, i.e. a bare color-neutral quark-antiquark pair.
To a nucleus with which it interacts during some finite time (of the order of a few fermi),
it appears generally speaking as a more complicated quantum state. When the onium has a large
rapidity with respect to the nucleus, its typical Fock state as ``seen'' from the point of view
of the nucleus is a dense set of gluons.

\begin{figure}[t]
  \centering
  \includegraphics[width=1.\textwidth,clip]{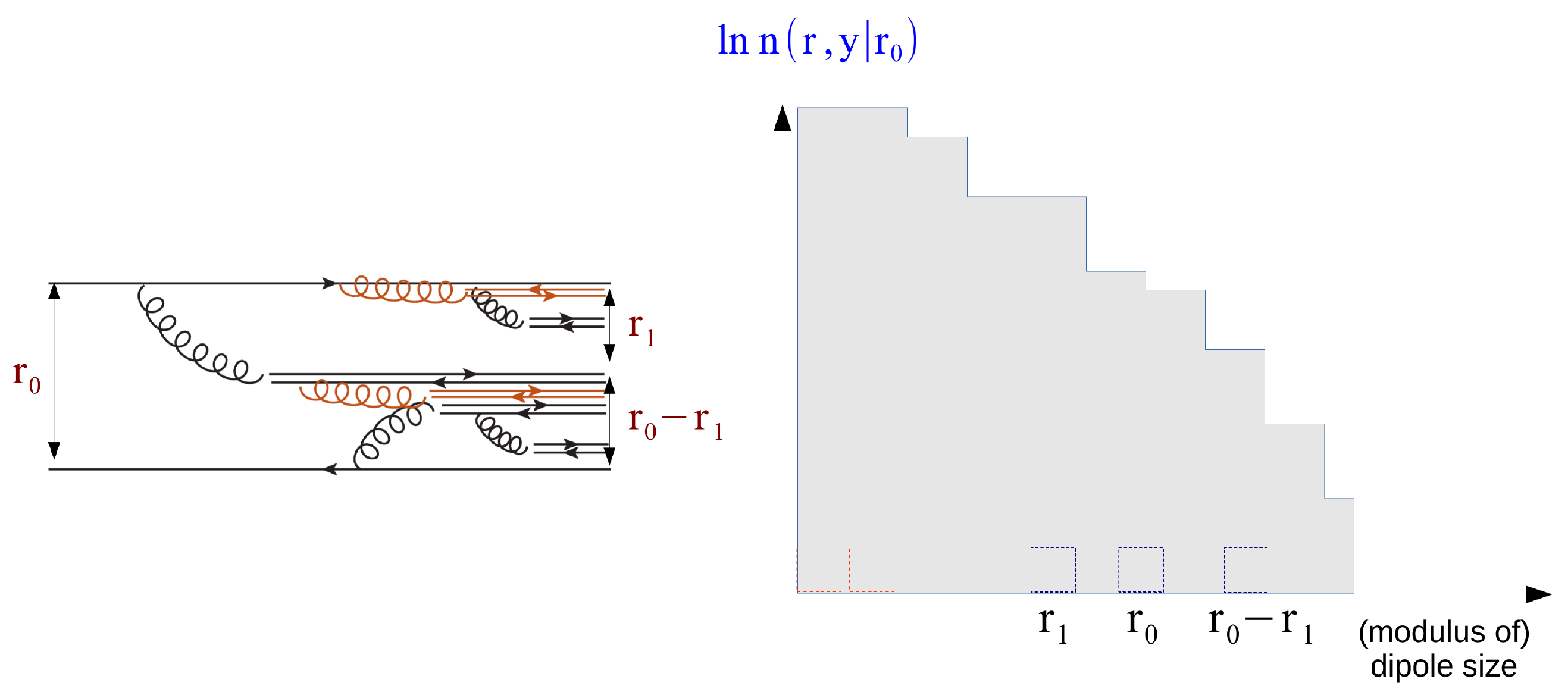}
  \caption{Schematic representation of QCD evolution in rapidity.
    {\it Left}: An initial $q\bar q$ pair of size $r_0$
    boosted to the rapidity $Y$ gets dressed
    with (essentially) gluons. In the color dipole model, in which color-neutral sets of gluons are
    represented by sets of color dipoles,
    the Fock state is built through a $1\rightarrow 2$
    branching process.
    {\it Right}: Plot of the dipole number $n$ at rapidity $Y$, in one particular realization
    of the QCD evolution of the onium, as a function of the dipole size. The mean shape is
    given by the solution to the deterministic BFKL equation, but fluctuations in the beginning
    of the evolution may shift the whole distribution randomly.
  }
\label{fig-1}
\end{figure}

The probability of a given partonic configuration can in principle be calculated from field theory.
But analytical expressions may be obtained only in appropriate asymptotic limits.
For large rapidities, the color dipole model~\cite{Mueller:1993rr} provides
a procedure to generate unweighted
parton configurations and to derive evolution equations in the rapidity variable.
It is formulated in the large-number-of-color $(N_c)$ limit of QCD,
in which a globally color-neutral set of gluons can be represented by color dipoles,
and in transverse position space.
In the framework of the dipole model, to an increase of the rapidity
of a color dipole by the small quantity $dy$ is associated a probability $dP$
that a gluon be emitted. If $r_0$ is the
size vector of the initial dipole,
then the probability that a gluon be emitted at position $r_1$ with respect to the quark,
up to $d^2r_1$,
reads~\cite{Mueller:1993rr}
\be
dP=\bar\alpha dy\frac{d^2r_1}{2\pi}\frac{r_0^2}{r_1^2(r_{0}-r_{1})^2},
\quad\text{with}\quad
\bar\alpha\equiv\frac{\alpha_s N_c}{\pi}.
\ee
In the large-$N_c$ limit, the emitted gluon together with the quark part of the initial dipole,
and the same gluon together with the antiquark part of the initial dipole,
form two new dipoles, which replace the initial one.
Furthermore, each dipole evolves independently
of the other ones present in the Fock state,\footnote{This is
  because the leading order in $1/N_c$ is given by
  planar graphs. Interference graphs for gluon emissions off different
  dipoles would be nonplanar, and hence can be discarded.}
until the maximum rapidity is reached; see Fig.~\ref{fig-1}.
Hence in the color dipole model, the state of the interacting onium
is built from a $1\rightarrow 2$ {\it branching process}
(for a review, see Ref.~\cite{Munier:2014bba}).

The mean number density of gluons, namely averaged over events,
is given by the well-understood solution to the
BFKL equation~\cite{Kovchegov:2012mbw}.
Event-by-event fluctuations are important in the initial stages of the evolution,
when the system is very dilute,
and at all rapidities only in the evolution of the dipoles which have sizes close to
that of the largest dipole in the event, for which the number density is low.
A complete description of these fluctuations is currently out of reach, but
we were able to model them and draw phenomenological consequences
that we shall now describe~\cite{Mueller:2014fba}.

%%%%%%%%%%%%

\section{Scattering cross sections}

Let us consider an onium of size~$r_0$, small compared to the typical
hadronic size $1/\Lambda_\text{QCD}$,
scattering off a large nucleus at total rapidity~$Y$.
We will discuss the total cross section.
From the point of view of the nucleus, at high total rapidity,
the latter typically appears dressed with many gluons, represented
by a set of color dipoles. The $S$-matrix element for the elastic scattering of
this particular state is the product of the $S$-matrix elements for the scattering of
the individual dipoles present in the onium at the time of its
interaction with the nucleus. If $\{r_i\}$ is the set of their size vectors, then
\be
\begin{split}
&S(r_0)=\prod_{i}
S_\text{MV}(r_i),\\
&\text{with}\quad S_\text{MV}(r)
=\exp\left[{-\frac{r^2Q_\text{MV}^2}{4}}\ln(e+4/r^2\Lambda_\text{QCD}^2)\right],
\end{split}
\ee
where $Q_\text{MV}\sim 1\ \text{GeV}$ is the nuclear saturation momentum. The expression for
the $S$-matrix element associated to the scattering of
an elementary dipole at low rapidity, $S_\text{MV}$, stems from
the McLerran-Venugopalan model~\cite{McLerran:1993ni}.

$S(r_0)$ is close to~1 if {\it no dipole} in the state has a size
much larger than $1/Q_\text{MV}$.
It is close to~0 if {\it at least one} dipole has a size larger than $1/Q_\text{MV}$.
In other words, in a first approximation,
$S(r_0)$ can be thought of as the probability that there be at least one dipole
larger than the inverse saturation momentum of the nucleus in the Fock state of the photon
at rapidity $Y$~\cite{Mueller:2014fba}. The amplitude for the scattering is just
$T(r_0,Y)=1-\langle S(r_0)\rangle_Y$, where
$\langle \cdot\rangle_Y$ represents the average over
all possible Fock states, namely dipole configurations.

$T(r_0,Y)$ also obeys the Balitsky-Kovchegov
equation~\cite{Balitsky:1995ub,Kovchegov:1999ua}, whose solution exhibits the following
feature:
\be
\begin{split}
&T(r_0,Y)\sim  \ln\frac{1}{r_0^2 Q_s^2(Y)}\left[r_0^2 Q_s^2(Y)\right]^{\gamma_0}\\
&\text{for}\quad 1\ll |\ln r_0^2Q_s^2(Y)|\ll \sqrt{\chi''(\gamma_0)\bar\alpha Y},
\end{split}
\ee
where the rapidity-dependent saturation scale $Q_s(Y)$ reads
\be
Q_s^2(Y)\sim Q_\text{MV}^2 \frac{e^{\bar\alpha\chi'(\gamma_0)Y}}{(\bar\alpha Y)^{3/(2\gamma_0)}}.
\label{eq:Qs}
\ee
$\chi(\gamma)=2\psi(1)-\psi(\gamma)-\psi(1-\gamma)$ and $\gamma_0$ solves
$\chi'(\gamma_0)=\chi(\gamma_0)/\gamma_0$.
The analytic form of the amplitude
can be related to the distribution of the two kinds of
fluctuations in the evolution of the
onium, the ones in the beginning of the evolution, and
the ones at the large-size tip of the distribution of the dipole sizes;
see Ref.~\cite{Mueller:2014fba} for details.

We now turn to the scattering of two small onia, of respective
sizes $r_0$ and $r_0'>r_0$. At moderate rapidities,
the amplitude solves the BFKL equation, but to date, there
is no sound formulation of the latter in the limit of large rapidities.
However, the picture of QCD evolution as a branching process enabled us to find
the asymptotic form of the amplitude~\cite{Mueller:2014fba}:
\be
T(r_0,Y)\sim \ln^2\frac{1}{r_0^2 Q_s^2(Y)}\left[r_0^2 Q_s^2(Y)\right]^{\gamma_0},
\ee
where now, $Q_s^2(Y)\simeq e^{\chi'(\gamma_0)\bar\alpha Y}/r_0'^2$.

Note that in this picture, the cross section is essentially determined by
the size distribution
of the {\it largest} dipole in the Fock state.

%%%%%%%%%%%%

\section{Fluctuations of the number of particles produced in onium-nucleus collisions}

Now consider the scattering of an onium of size {\it larger} than $1/Q_\text{MV}$ off a large
nucleus. When the impact parameter is within the radius of the nucleus, then
the $S$-matrix element is clearly close to zero.
One interesting quantity to study is the number of particles produced in a given
rapidity slice in the fragmentation region of the proton {\it in a given event}.
In Ref.~\cite{Liou:2016mfr}, we suggested that it may be directly related to the gluon density
in the onium integrated up to the saturation scale of the nucleus with which
it interacts at a value of the momentum fraction~$x$ in correspondence with the
rapidity of the final-state hadrons that are measured.
Hence measurements of the multiplicity of particles
may provide an insight into the {\it event-by-event fluctuations}
of the gluon density in the onium.

The dipole model, supplemented with a model for confinement,
is an appropriate framework to compute these fluctuations: Indeed,
it provides equations for the probability $P_n$ to have $n$ dipoles of size larger
than a given threshold, namely
$n$ gluons of momenta smaller than the saturation scale, in a given realization.
But the equations for $P_n$ are too complicated to be solved analytically.
We were however able to obtain closed expressions for $P_n$ in asymptotical limits.
For $n$ large with respect to its expected value~$\bar n$,
we have found~\cite{Liou:2016mfr}
\be
P_n\propto r_0^2\Lambda_\text{QCD}^2\times\frac{1}{n_1}e^{-n/n_1}
\quad\text{for}\ {n\gg\bar n}.
\label{eq:Pn_large}
\ee
The mechanism for generating high-multiplicity events is the following:
In the very beginning of the evolution, the onium of size $r_0$ promptly splits
to the largest dipoles compatible with confinement,
of size $1/\Lambda_\text{QCD}$. This step has the probability
$r_0^2\Lambda_\text{QCD}^2$ in the limit $r_0\ll 1/\Lambda_\text{QCD}$.
In a second step, the system decays into smaller dipoles over the remaining rapidity range:
The number of dipoles resulting from this process
is distributed as a decreasing exponential, in the same way as the
number of objects generated in a $1\rightarrow 2$ simple branching process.
The parameter $n_1$ of the exponential
is the expected number of dipoles, namely the ordinary gluon density in a hadron
evaluated at the saturation scale of the nucleus.

In the small-$n$ limit instead, we have found the following expression
for~$P_n$~\cite{Domine:2018myf}:
\be
P_n\propto \exp\left(\frac12\ln^2n-\frac14\frac{\ln^4 n}{\ln^2\bar n}
-\frac14\ln^2\bar n\right)\quad \text{for}\ {1\ll n\ll\bar n}.
\label{eq:Pn_small}
\ee

In order to check our understanding of the mechanism leading
to dipole number fluctuations,
we have written a Monte Carlo code implementing
the color dipole model
(supplemented with an infrared cutoff modeling confinement)~\cite{Domine:2018myf}.
The numerical results we have obtained are in very good agreement with the
analytical formulas~(\ref{eq:Pn_large},\ref{eq:Pn_small}).

%%%%%%%%%%%%%%

\section{Outlook}

From our recent theoretical results on the multiplicity distribution in the final state
of onium-nucleus scattering processes, one may derive predictions for proton-nucleus collisions
at the LHC, or for deep-inelastic scattering at a future Electron-Ion Collider.
The former is quite demanding, since a proton is not a quark-antiquark pair, and it is
not clear whether a quark-diquark dipole would be a good model in this
context. The latter is clearly more straightforward~\cite{giacalone_to-appear},
since onium-nucleus interaction is a subprocess of electron-nucleus scattering,
factorizable in the limit of high energies.

Next-to-leading order effects are known to be important practically; one
eventually needs to implement them to arrive at a quantitative description of the data.
A kinematical constraint~\cite{Motyka:2009gi}, or a modified kernel incorporating
resummations of collinear
logarithms~\cite{Iancu:2015vea} may be good ways to effectively account for
subleading effects.

%%%%%%%%%%%%%%

\section*{Acknowledgements}

We thank Victor P. Gon\c{c}alves, Leszek Motyka and Michael J. Murray for stimulating feedback.
This work was supported in part by the Agence Nationale de la Recherche under the project \# ANR-16-CE31-0019.

%\bibliographystyle{woc}
%\bibliography{biblio_proceedings}

\begin{thebibliography}{12}

\bibitem{Mueller:1993rr}
A.H. Mueller, Nucl. Phys. \textbf{B415}, 373 (1994)

\bibitem{Munier:2014bba}
S.~Munier, Sci. China Phys. Mech. Astron. \textbf{58}, 81001 (2015),
  \texttt{1410.6478}

\bibitem{Kovchegov:2012mbw}
Y.V. Kovchegov, E.~Levin, \emph{{Quantum chromodynamics at high energy}}
  (Cambridge University Press, 2012), ISBN 978-0521112574

\bibitem{Mueller:2014fba}
A.H. Mueller, S.~Munier, Phys. Lett. \textbf{B737}, 303 (2014),
  \texttt{1405.3131}

\bibitem{McLerran:1993ni}
L.D. McLerran, R.~Venugopalan, Phys. Rev. \textbf{D49}, 2233 (1994),
  \texttt{hep-ph/9309289}

\bibitem{Balitsky:1995ub}
I.~Balitsky, Nucl. Phys. \textbf{B463}, 99 (1996), \texttt{hep-ph/9509348}

\bibitem{Kovchegov:1999ua}
Y.V. Kovchegov, Phys. Rev. \textbf{D61}, 074018 (2000), \texttt{hep-ph/9905214}

\bibitem{Liou:2016mfr}
T.~Liou, A.H. Mueller, S.~Munier, Phys. Rev. \textbf{D95}, 014001 (2017),
  \texttt{1608.00852}

\bibitem{Domine:2018myf}
L.~Domin\'e, G.~Giacalone, C.~Lorc\'e, S.~Munier, S.~Pekar (2018),
  \texttt{1810.05049}

\bibitem{giacalone_to-appear}
G.~Giacalone, S.~Munier, {work in progress}

\bibitem{Motyka:2009gi}
L.~Motyka, A.M. Stasto, Phys. Rev. \textbf{D79}, 085016 (2009),
  \texttt{0901.4949}

\bibitem{Iancu:2015vea}
E.~Iancu, J.D. Madrigal, A.H. Mueller, G.~Soyez, D.N. Triantafyllopoulos, Phys.
  Lett. \textbf{B744}, 293 (2015), \texttt{1502.05642}

\end{thebibliography}

\end{document}